\documentclass[twocolumn]{aastex63}
\usepackage{hyperref}
\usepackage[colorinlistoftodos]{todonotes}
\usepackage{amsmath}
\usepackage{nicefrac}
\shorttitle{Galaxy Dust Maps}
\usepackage[encapsulated]{CJK}

\newcommand{\microns}{$\mu$m}
\newcommand{\sersic}{S\'{e}rsic}
\newcommand{\scarlettwo}{{\sc scarlet\oldstylenums{2}}}

\begin{document}

\title{Spatially Resolved Galaxy--Dust Modeling with Coupled Data-Driven Priors}

\correspondingauthor{Jared Siegel}
\email{siegeljc@princeton.edu}

\author[0000-0002-9337-0902]{Jared C. Siegel}
\altaffiliation{NSF Graduate Research Fellow}
\affiliation{Department of Astrophysical Sciences, Princeton University, Princeton, NJ 08544, USA}

\author[0000-0002-8873-5065]{Peter Melchior}
\affiliation{Department of Astrophysical Sciences, Princeton University, Princeton, NJ 08544, USA}
\affiliation{Center for Statistics and Machine Learning, Princeton University, Princeton, NJ 08544, USA}

\begin{abstract}
    A notorious problem in astronomy is the recovery of the true shape and spectral energy distribution (SED) of a galaxy despite attenuation by interstellar dust embedded in the same galaxy.
    This problem has been solved for a few hundred nearby galaxies with exquisite data coverage, but these techniques are not scalable to the billions of galaxies soon to be observed by large wide-field surveys like LSST, Euclid, and Roman.
    We present a method for jointly modeling the spatially resolved stellar and dust properties of galaxies from multi-band images.
    To capture the diverse geometries of galaxies, we consider non-parametric morphologies, stabilized by two neural networks that act as data-driven priors:
    the first informs our inference of the galaxy's underlying morphology, the second constrains the galaxy's dust morphology conditioned on our current estimate of the galaxy morphology.
    We demonstrate with realistic simulations that we can recover galaxy host and dust properties over a wide range of attenuation levels and geometries.
    We successfully apply our joint galaxy--dust model to three local galaxies observed by SDSS.
    In addition to improving estimates of unattenuated galaxy SEDs, our inferred dust maps will facilitate the study of dust production, transport, and destruction.
\end{abstract}


\section{Introduction}
\label{sec:intro}

Despite their small size ($\lesssim1$~{\microns} diameter), interstellar dust grains loom large in the study of galaxies.
They facilitate the formation of stars \citep[e.g.,][]{Hirashita2002} and are key to the heating and cooling of the interstellar medium \citep[ISM,][]{Draine2003}.
Interactions between photons and dust grains also reshape the  spectrum of starlight emitted by a galaxy.
Along a single line of sight, photons can be absorbed or scattered by dust grains (dust extinction).
When observing galaxies (or regions of galaxies), we must consider both extinction and the possibility of dust grains scattering photons into the line of sight \citep[dust attenuation,][]{Salim2020}.
Roughly a third of stellar light is reprocessed into the infrared by dust grains \citep[e.g.,][]{Bernstein2002}.
By characterizing a galaxy's dust content, we can probe the inner workings of the ISM and recover the emitted stellar spectrum.

For hundreds of local galaxies, the spatial distribution of dust has been mapped with a variety of tracers, e.g., the SDSS-IV MaNGA IFU survey \citep[e.g.,][]{Bundy2015,Greener2020} and the DustPedia database of archival GALEX, SDSS, DSS, 2MASS, WISE, Spitzer, and Planck observations \citep[e.g.,][]{Davies2017, Casasola2017}.
However, galaxies with such comprehensive observations are rare.
For the vast majority of observed galaxies, we are limited to multi-band optical and near-infrared (NIR) images.
In this case, the only tracer of dust attenuation is the shape of the galaxy's spectral energy distribution (SED); dust attenuation more efficiently suppresses blue photons, resulting in redder SEDs in regions with more dust.

The physical properties of a galaxy are encoded in its SED, including mass, redshift, and dust reddening \citep[e.g.,][]{Conroy2013}.
Current methods typically parameterize an intrinsic SED and assume a uniform foreground dust screen.
In reality, dust attenuation is often nonuniform within a galaxy \citep[e.g.,][]{Davies2017}, resulting in biases on the inferred mass and redshift \citep[][]{Hahn2024}.
To overcome these biases, some studies allow for spatial variation in dust attenuation \citep[e.g,][]{Wuyts2012,Suess2019}.
Binning methods often assume a simplified galaxy morphology (e.g., concentric elliptical annuli) or  partition the galaxy based on a target per-bin signal-to-noise ratio \citep[e.g., Voronoi binning,][]{Cappellari2003}.
However, these methods lower our spatial resolution and do not necessarily divide galaxies into homogeneous regions.

\begin{figure*}[t!]
\gridline{\fig{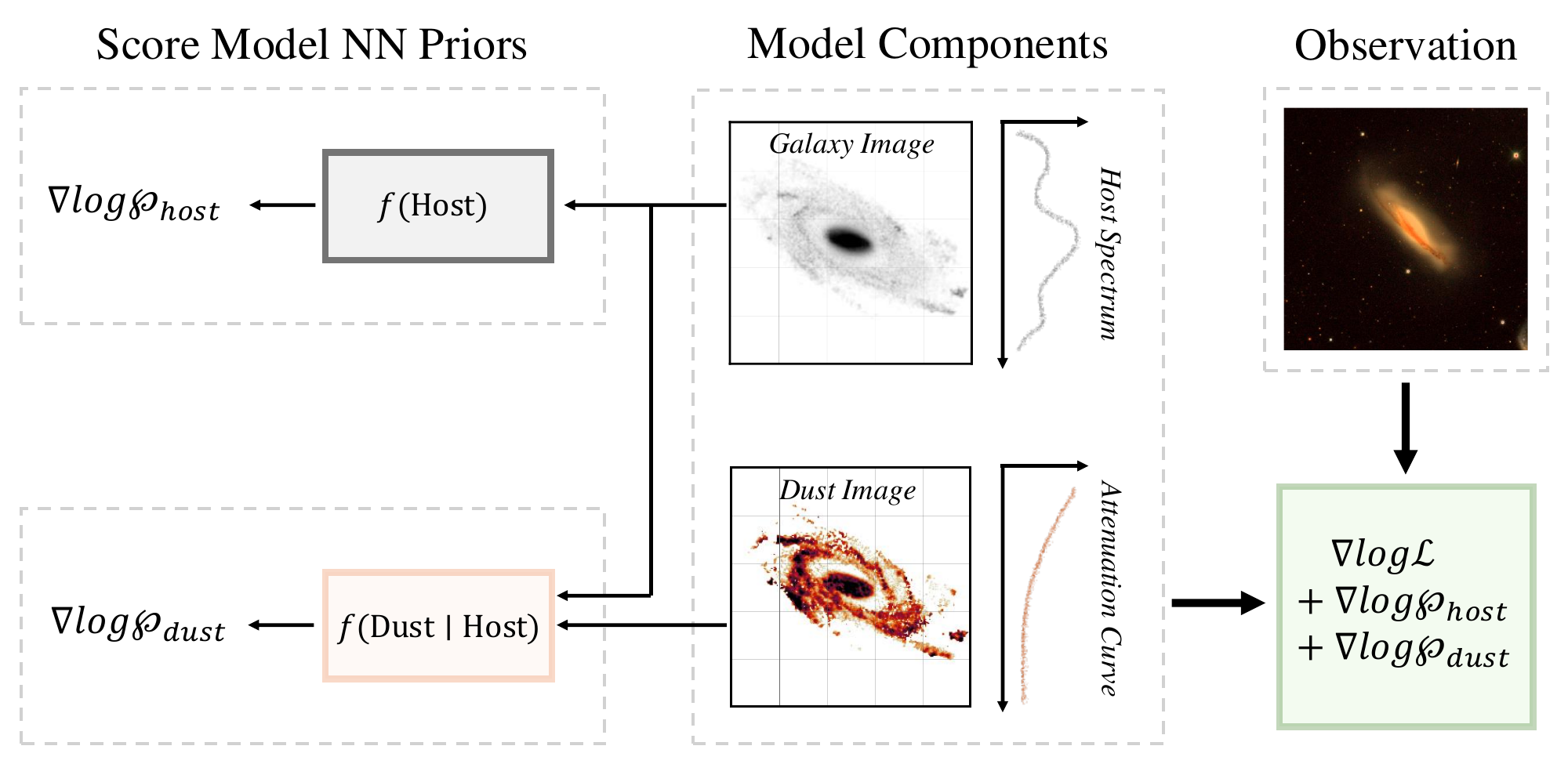}{
\textwidth}{ } }
\caption{
Our method for modeling a dust reddened galaxy.
Data-driven neural network priors inform our gradient-based inference of the galaxy and dust attenuation morphologies.
The dust prior is conditional on the host galaxy morphology.
}
\label{fig:visual_abstract}
\end{figure*}

On-going and upcoming wide-field surveys will soon observe billions of galaxies in the optical and NIR: the Vera C. Rubin Observatory’s Legacy Survey of Space and Time \citep[LSST,][]{LSST2009}, the ESA Euclid satellite mission \citep{Laureijs2011}, and the Nancy Grace Roman Space Telescope \citep{Spergel2015}.
With improved treatment of dust attenuation, our understanding of these galaxies will be more accurate and nuanced.
To prepare for these large surveys, we introduce a framework for recovering a galaxy's spatially resolved stellar component and dust properties from multi-band images.
We consider non-parametric galaxy and dust morphologies and utilize two data-driven priors: one for the galaxy's shape and one for the dust morphology (conditioned on the galaxy's shape); see \autoref{fig:visual_abstract} for a schematic.
To limit confounding variables, we do not include stellar population synthesis in our joint galaxy--dust fits; when fitting a multi-band image, we only require the inferred galaxy spectrum be positive.

We outline our method in \autoref{sec:methods}.
Our results, from both simulated and real galaxies, are presented in \autoref{sec:results}.
We discuss our findings in \autoref{sec:disc} and conclude in \autoref{sec:conclusions}.

\section{Methods}
\label{sec:methods}

\subsection{Joint galaxy and dust model}

Dust attenuation reduces the emitted intensity of light $F_\mathrm{host}(\lambda)$ to the observed intensity $F(\lambda)$, which is commonly expressed as
\begin{equation}
    \label{eqn:attenuation law}
   F(\lambda) = F_\mathrm{host}(\lambda)\, 10^{-0.4 A(\lambda)  },
\end{equation}
where $A(\lambda)$ is the wavelength dependence of dust attenuation.
We adopt the empirical relation of \citet{Calzetti2000} with modifications by \citet{Noll2009-cr} and \citet{Kriek2013-it},
\begin{align}
    A(\lambda) = A_0 k(\lambda, \delta),
\end{align}
where $A_0$ sets the attenuation amplitude at a reference wavelength $\lambda_0$, and $k(\lambda, \delta)$ is a polynomial with a single free parameter $\delta$ that modifies the slope of the attenuation curve.
It is common to express the level of attenuation in terms of V--band ($\lambda_0 \approx 5510~\mathrm{\AA}$).

For observed multi-band images, the spatial and spectral variation of the galaxy are integrated and discretized. If we stack these images and assume that spatial and spectral variation are independent, we can model the image cube as
\begin{equation}
    \label{eqn:host}
    \mathbf{Y} = \mathbf{F}^\mathrm{T} \times \mathbf{S},
\end{equation}
where $\mathbf{F}= (\int \mathrm{d}\lambda\,F(\lambda) c_i(\lambda),\dots)$ is the vectorized spectrum for a list of filter curves $c_i(\lambda)$, and $\mathbf{S}$ is a monochrome image of the  intensity variation (morphology).
$\mathbf{F}$ is one-dimensional, with a length equal to the number of observed filters $N_\mathrm{band}$; $\mathbf{S}$ is two-dimensional ($64\times64$); $\mathbf{Y}$ is three-dimensional  ($N_\mathrm{band} \times 64 \times 64$).
This parameterization has been found to be quite effective for galaxy images \citep{Melchior2018-hb}, but it does not account for color variations within a galaxy (e.g., dust attenuation or age gradients).
With dust, \autoref{eqn:host} needs to be modified according to \autoref{eqn:attenuation law}:
\begin{equation}
    \label{equ:ext_gal}
    \mathbf{Y} = (\mathbf{F}^\mathrm{T} \times \mathbf{S}) \odot 10^{-0.4 \mathbf{A} },
\end{equation}
where $\mathbf{A}$ denotes the multi-band image cube of dust attenuation. 
If we make the same assumption of spatial--spectral separability (i.e., $A_V$ is free to vary across the galaxy but $\delta$ is spatially uniform), we can express the attenuation cube as
\begin{equation}
    \label{eqn:a_cube}
    \mathbf{A} = \mathbf{K}^\mathrm{T} \times [-2.5 \log_{10}(\mathbf{D})     ],
\end{equation}
where $\mathbf{K} = (k(\lambda_i, \delta), \dots )$ is the attenuation curve evaluated for filters $i$, and  $\mathbf{D}$ is the image of the spatial distribution of $\frac{F}{ F_\mathrm{host}}(\lambda_0)$.
$\mathbf{K}$ is one-dimensional, with length $N_\mathrm{band}$, $\mathbf{D}$ is two-dimensional ($64 \times 64)$, and $\mathbf{A}$ is three-dimensional ($N_\mathrm{band} \times 64 \times 64)$.

\subsection{Data-driven priors}

Fitting a joint galaxy--dust model to multi-band observations faces severe degeneracies.
The presence of dust is revealed as a reddening of the galaxy's light, but in areas without light, any amount of attenuation is consistent with the data.
Additionally, the likelihood will favor dust attenuation for any reddened pixels, thereby over fitting the data. 
As a result, the likelihood alone cannot distinguish between physical and non-physical galaxy or dust morphologies.
This inverse problem thus requires priors for at least the galaxy morphology $\mathbf{S}$ and the dust morphology $\mathbf{D}$.
Due to the complications of stellar population synthesis, we do not regularize the host spectrum parameter with an informative prior; we only require $\mathbf{F}>0$.
Future implementations may want to represent the host spectrum with a stellar population synthesis model (\autoref{sec:disc}).

We expect dust reddening and starlight to be correlated; reddening naturally cannot be measured in regions without galaxy light, and the efficiency of dust production/destruction is known to depend on   the local environment.
We therefore posit a dependency structure of the prior in the form of
\begin{equation}
\label{eqn:prior_structure}
p(\mathbf{S},\mathbf{D}) = p(\mathbf{S})\,p(\mathbf{D}\mid\mathbf{S}).
\end{equation}

We approximate \autoref{eqn:prior_structure} with neural networks, which have been shown to accurately represent prior distributions \citep[e.g.,][]{Van_den_Oord2016-mc}.
Score-matching models, which underpin diffusion models \citep{Song2019,Song2020}, are particularly effective when considering high-dimensional parameters.
Given a dataset $\{\mathbf{x}_i\}$ of independent samples from a data distribution $p_\mathrm{data}(\mathbf{x})$, a network $s_{\boldsymbol{\theta}}(\mathbf{x})$ with parameters $\boldsymbol{\theta}$ is trained to approximate the score function $\nabla \log p_\mathrm{data}(\mathbf{x})$.
In our case,  $p_\mathrm{data}(\mathbf{x})$ is the distribution of galaxy or dust morphologies.
By evaluating the gradient $\nabla \log p_\mathrm{data}(\mathbf{x})$ in each iteration of our joint galaxy--dust fit, we can encourage morphologies that are consistent with $p_\mathrm{data}(\mathbf{x})$.

We train our priors following \cite{Song2020}, as briefly outlined below.
We perturb each draw from $p_\mathrm{data}(\mathbf{x})$ by a Gaussian noise distribution: $\tilde{\mathbf{x}}_i = \mathbf{x}_i + \mathcal{N}(\mathbf{x}_i \mid \sigma^2 \mathbf{I})$; the perturbed distribution is denoted $q(\tilde{\mathbf{x}} \mid \mathbf{x}, \sigma)$.
The network that minimizes the likelihood function 
\begin{align}
    l(\theta \mid \sigma) = \mathbb{E}_{ \{ \mathbf{x} \} } \mathbb{E}_{ \mathbf{\tilde{x}}  \sim  \mathcal{N}(\mathbf{x}, \sigma^2 \mathbf{I}) }  & \Big [  \big \| \mathbf{s}_\theta ( \mathbf{\tilde{x}} \mid \sigma )  \nonumber \\ 
     -  \nabla \log & q(\tilde{\mathbf{x}} \mid \mathbf{x}, \sigma) \big \|^2_2 \Big ]
\end{align}
is known to satisfy $s_\theta(\tilde{\mathbf{x}} \mid \sigma) = \nabla \log q (\tilde{\mathbf{x}} \mid \sigma)$, where $q (\tilde{\mathbf{x}} \mid \sigma) = \int q(\tilde{\mathbf{x}} \mid \mathbf{x}, \sigma) p_\mathrm{data}(\mathbf{x}) d \mathbf{x}$.
For a Gaussian noise distribution, $\nabla \log q(\tilde{\mathbf{x}} \mid \mathbf{x}, \sigma) = -( \tilde{\mathbf{x}} - \mathbf{x})/ \sigma^2$.
At low noise levels---i.e., $q(\tilde{\mathbf{x}} \mid \sigma) \approx p_\mathrm{data}(\tilde{\mathbf{x}})$---it follows $s_\theta(\tilde{\mathbf{x}} \mid \sigma) = \nabla \log q (\tilde{\mathbf{x}} \mid \sigma) \approx \nabla \log p_\mathrm{data} (\tilde{\mathbf{x}} \mid \sigma)$.

\begin{figure*}[t!]
\gridline{\fig{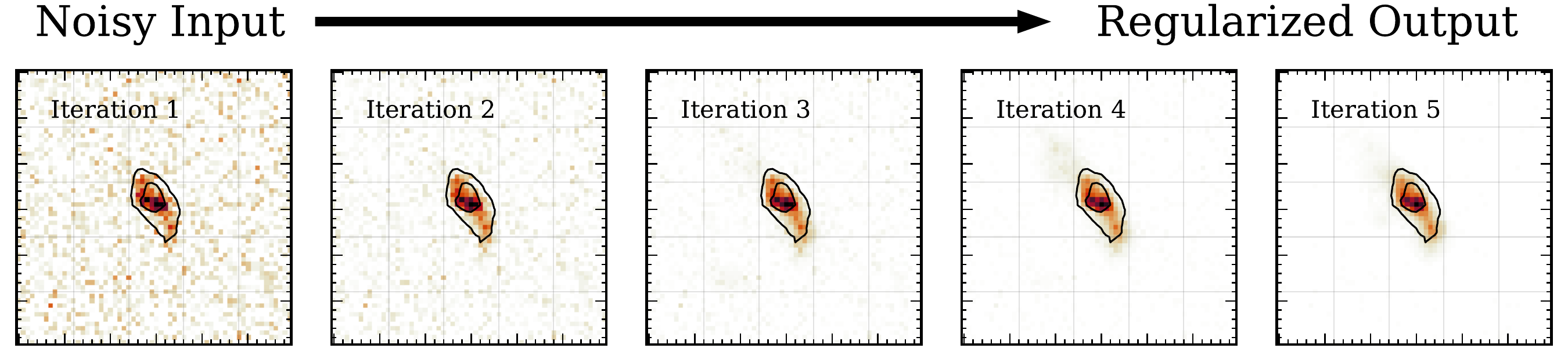}{
\textwidth}{ } }
\caption{
An initially noisy dust distribution is transformed into a plausible dust distribution by iteratively applying our score-matching prior (left to right).
The color scale covers per-pixel $A_V \in [0, 1.25]$. 
The contours present the $50$ and $84$-th percentiles of the unreddened galaxy light.
}
\label{fig:prior_regularization}
\end{figure*}

For the host morphologies, we adopt the publicly available score-matching network of \cite{Sampson2024}.\footnote{\url{https://github.com/SampsonML/galaxygrad}}
That model was trained on 600,000 galaxy images from the Subaru Hyper-Suprime Cam catalog \citep{Bosch2018}.
To calculate the prior gradient for a given morphology $\mathbf{S}$, we ideally would evaluate the score model at $\sigma=0$.
However, the finite size of the training set leads to unstable behavior as $\sigma \rightarrow 0$. 
In this work, we use $\sigma = 0.02$. 
Compared to heuristic constraints (e.g., requiring a symmetric and centrally peaked morphology), \cite{Sampson2024} demonstrated that the data-driven prior significantly improves the recovery of galaxy spectra and morphologies. 

We build the dust morphology prior as a score-matching model that is conditioned on the host morphology: $p(\mathbf{D} \mid \mathbf{S})$.
We adopt a U-Net architecture for the neural network \citep{Ronneberger2015}.
The network takes a current estimate of the host and dust morphologies ($2 \times 64 \times 64$) and outputs the gradient of the logarithmic dust distribution prior ($64 \times 64$).
We train the network with simulated galaxies from \citet{Faucher2023}.
The simulated galaxies are drawn from the NIHAO project's redshift zero snapshots \citep{Wang2015}, which include hydrogen, helium, and metal-line cooling \citep[][]{Shen2010} and stellar feedback \citep[][]{Stinson2013}.
Mock images, either with or without the effects of dust, were generated with SKIRT, a Monte Carlo radiative transfer code \citep{Camps2020}.
The \citet{Faucher2023} galaxies successfully reproduce the color distribution of the observed DustPedia sample  \citep[$>800$ local galaxies,][]{Davies2017}.
Unlike observed objects, these simulations allow us to calculate the ground-truth host and dust attenuation morphologies.
Multi-band image cubes are publicly available for 65 galaxies, with 10 viewing angles for each.\footnote{\url{https://github.com/ntf229/NIHAO-SKIRT-Catalog}}
Such a low number of training samples is only acceptable because the dependency structure---i.e., the dust distribution given the light distribution---is highly informative.
Due to the smaller training sample, we evaluate the dust score model at $\sigma = 0.1$.

The behavior of our data-driven dust morphology prior is demonstrated in \autoref{fig:prior_regularization}.
Starting with an initially noisy dust morphology, we repeatedly update the image by following the positive gradient of the prior from the score model, yielding a reasonable dust morphology.
This regularization is vital to non-parametric galaxy--dust modeling.

\subsection{Model optimization}
\label{sec:optimization}

We have added the joint galaxy--dust model (\autoref{equ:ext_gal} and \autoref{eqn:a_cube}) to the \scarlettwo{}\footnote{\url{https://github.com/pmelchior/scarlet2}} modeling framework \citep[][ Melchior et al., in prep.]{Sampson2024}, which is implemented in \texttt{jax}\ \citep{jax2018} and \texttt{equinox} \citep{equinox2021}.
We run gradient-descent with the Adam optimizer \citep{Kingma2015-pq} to determine the maximum a posteriori (MAP) estimates of our galaxy--dust model.
The free parameters are the galaxy's intrinsic spectrum ($\mathbf{F}$) and morphology ($\mathbf{S}$) as well as the attenuation law (parameterized by the power-law index $\mathbf{\delta}$) and morphology ($\mathbf{D}$) of the dust.
The galaxy and dust morphologies are subject to the data-driven priors outlined above.
We adopt a uniform prior on the attenuation power-law index $\delta \in [-1,0.5]$.
To limit complexity, we only require the host spectrum obey $\mathbf{F}>0$; a possible extension is to treat the host spectrum as a stellar population synthesis model (see \autoref{sec:disc} for more discussion of this aspect).
The optimization is terminated based on iteration-to-iteration convergence of the host spectrum parameter, the most stable global parameter in this problem.

The model is initialized based on a dust-free fit to the galaxy. 
We begin with near zero dust attenuation and set the galaxy spectrum to the observed (i.e., dust reddened) values.
Starting with low dust attenuation is critical, due to the potential degeneracies in our joint galaxy--dust model.
If our model enters the thin-screen limit (i.e., the entire galaxy is covered in dust), increasing the level of attenuation is completely degenerate with making the host spectrum bluer, because we have no prior on the host spectrum other than positivity.
Once in the thin-screen limit, the likelihood cannot move the model towards lower dust levels, and the dust morphology prior is unlikely to do so, because many galaxies in the training set have spatially extended dust morphologies.
Treating the host spectrum as a stellar population synthesis model would help alleviate the thin-screen degeneracy.

\begin{figure*}[t!]
\gridline{\fig{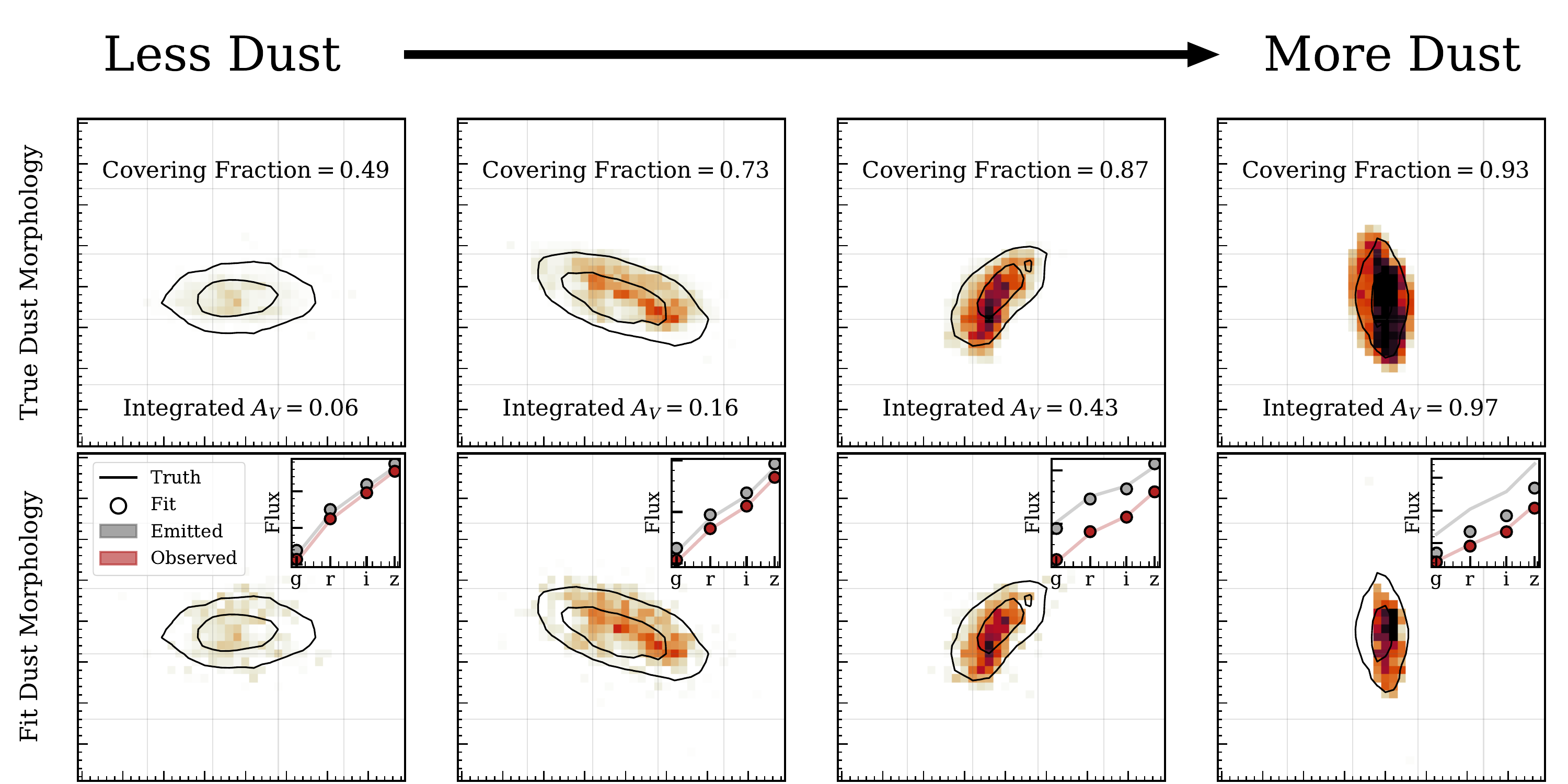}{
\textwidth}{ } }
\caption{
Ground truth (\emph{top}) and inferred (\emph{bottom}) dust morphologies for four representative simulated galaxies.
The color scale covers per-pixel $A_V\in[0, 1.25]$.
The contours present the $50$ and $84$-th percentiles of the unreddened galaxy light.
The inset panels show the spectrum of the galaxy before (grey) and after (red) dust attenuation.
We successfully recover the amplitude and morphology of dust attenuation for a wide range of reddening levels.
}
\label{fig:dust_fit_grid}
\end{figure*}

Depending on what filters are available for our multi-band observation,
we consider two scenarios for initializing the host galaxy's intrinsic morphology.
If the source is observed at wavelengths where dust attenuation is negligible ($\lambda \gtrsim 1$~{\microns}), the host's intrinsic morphology can easily be inferred from the reddest image.
In this case, we initialize the host morphology according to the reddest available band. 
We refer to this scenario as the \textit{red} start.

If multi-band observations do not include sufficiently red filters, only reddened versions of the host's  morphology are visible.
We refer to this scenario as the \textit{blue} start.
In this case, we initialize the host's intrinsic morphology based on a rough estimate of the galaxy's dust morphology.
We first calculate a color for all pixels within the galaxy footprint.
We then approximate the intrinsic host spectrum by co-adding the bluest $N\%$ of the pixels; here we set $N=50\%$.
For each of the red pixels, we calculate the level of dust reddening relative to our estimated host spectrum.
The resulting dust map $\mathbf{A}$ is then used to estimate the host's intrinsic morphology $\mathbf{S}$ from the observed dust reddened morphology $\mathbf{S}_\mathrm{obs}$: 
\begin{equation}
    \mathbf{S} = \mathbf{S}_\mathrm{obs} / 10^{-0.4 \mathbf{A}}.
\end{equation}
If the signal-to-noise level is high, dust is the dominant source of color gradients, and parts of the galaxy are unreddened, this process accurately estimates the dust morphology (in the limit $N\rightarrow0\%$).
For non-zero noise levels, this method yields noisy dust maps, which are only acceptable thanks to the regularization from our priors.

\section{Results}
\label{sec:results}

\begin{figure*}[t!]
\gridline{\fig{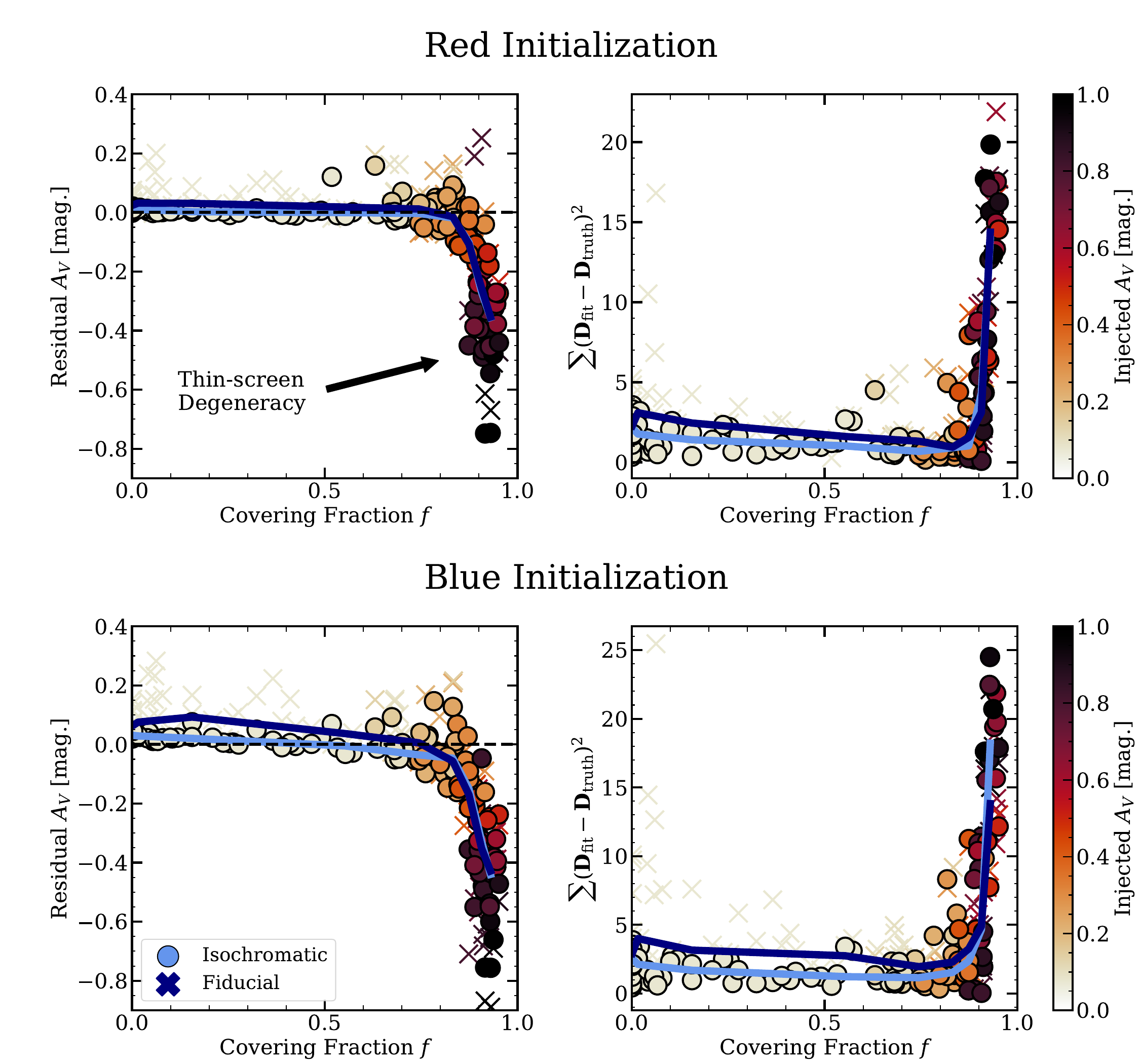}{
\textwidth}{ } }
\caption{
Accuracy of our inferred attenuation level (\emph{left}) and the difference between the inferred dust map and the true dust distribution (\emph{right}) for 65 simulated galaxies with three orientation angles each.
The solid lines represent a running median, for either the isochromatic or fiducial galaxies.
In the top row, we assume that we have a good estimate of the unreddened galaxy morphology from a NIR image; in the bottom, we restrict our initialization to optical bands.
Marker color represents ground-truth $A_V$.
Our joint galaxy--dust modeling is robust for dust covering fractions $f\lesssim 0.9$ (see \autoref{sec:disc}).
}
\label{fig:dust_fit_summary_breakdown}
\end{figure*}

We apply our joint galaxy--dust model to $griz$ multi-band images of simulated \citep{Faucher2023} and observed \citep[SDSS,][]{York2000} galaxies.
In our current implementation, we assume the dust attenuation curve and the galaxy's emitted SED are near-uniform within a given filter.
We therefore omit $u$~band because of the steepness of the attenuation curve in the near-UV.

\subsection{Simulated galaxies}

We begin by considering 65 simulated galaxies from  \cite{Faucher2023}.
For each galaxy, we fit synthetic $griz$ observations, with per-band S/N$=50$.

Our fits to four representative galaxies are presented in \autoref{fig:dust_fit_grid}. 
For each galaxy, we report the inferred dust maps and host SED, alongside the ground-truth dust and host properties.
Reflecting a scenario where we have NIR filter coverage, we initialize the fits with the unreddened galaxy morphologies; 
we revisit this assumption below.
For a wide range of attenuation levels, our joint galaxy--dust model successfully determines the dust and host properties.
In particular, we recover accurate dust attenuation maps and dereddened host SEDs.
However, when the entire galaxy is covered by dust (i.e., a thin-screen), there is a degeneracy between the host's emitted spectrum and the level of dust attenuation, because we apply no prior on the host spectrum other than positivity.
We discuss remedies, such as representing the host spectrum with physically motivated templates, in \autoref{sec:disc}.
Outside the thin-screen limit, our modeling successfully recovers the host's SED and the dust morphology.

To establish the impact of the thin-screen degeneracy on our current implementation, we consider all 65 galaxies from \cite{Faucher2023} and three orientation angles for each.
In \autoref{fig:dust_fit_summary_breakdown}, we compare the recovered galaxy and dust properties with ground truth.
For covering fractions $f<90\%$, where $f$ is the fraction of the emitted light with non-zero attenuation, we accurately infer the amplitude and morphology of dust attenuation. 
At higher covering fractions, we are subject to the thin-screen degeneracy; in these cases, our model still captures the dust morphology of the most strongly reddened pixels but systematically underestimates the level of attenuation. 

Dust attenuation is not the only source of color variations across a galaxy.
Measurement noise creates apparent color variations in multi-band imaging; 
however, our data-driven prior strongly disfavors  dust morphologies without significant pixel-to-pixel correlations, thereby suppressing pixel noise (see \autoref{fig:prior_regularization}).
Color variations can also arise from the galaxy itself, most notably from differences in stellar age across the galaxy.
Since stellar age gradients can have similar spatial structures to dust morphologies, we may mistakenly interpret variations in the stellar composition that lead to redder emission as dust attenuation. 
The aid of our data-driven priors may somewhat reduce this misassociation but cannot fundamentally prevent it.

To quantify the impact of non-dust color variations, we create ``isochromatic" galaxies by replacing their spatially variable stellar emission with their average emission SED and redden it by the same dust map as the original galaxies.
For each simulated galaxy, we apply our joint galaxy--dust model to the isochromatic multi-band image cube.
The recovered galaxy and dust properties are summarized in \autoref{fig:dust_fit_summary_breakdown}. 
For the majority of galaxies, the modeling quality is similar between the fiducial and isochromatic galaxies.
Differences primarily arise for galaxies with very low attenuation ($A_V \lesssim 0.1$).
For these low attenuation galaxies, variation in age and metallicity of stellar populations are the dominant source of color variations;
this can result in overestimation of the reddening level, if dust is invoked to explain color variations in the stellar populations.
However, fits that overestimate the level of reddening are still rare; in \autoref{fig:dust_fit_summary_breakdown}, we see the median residual $A_V$ lies near zero for both the fiducial and monochromatic galaxies at $A_V<0.1$.

\begin{figure*}[t!]
\gridline{\fig{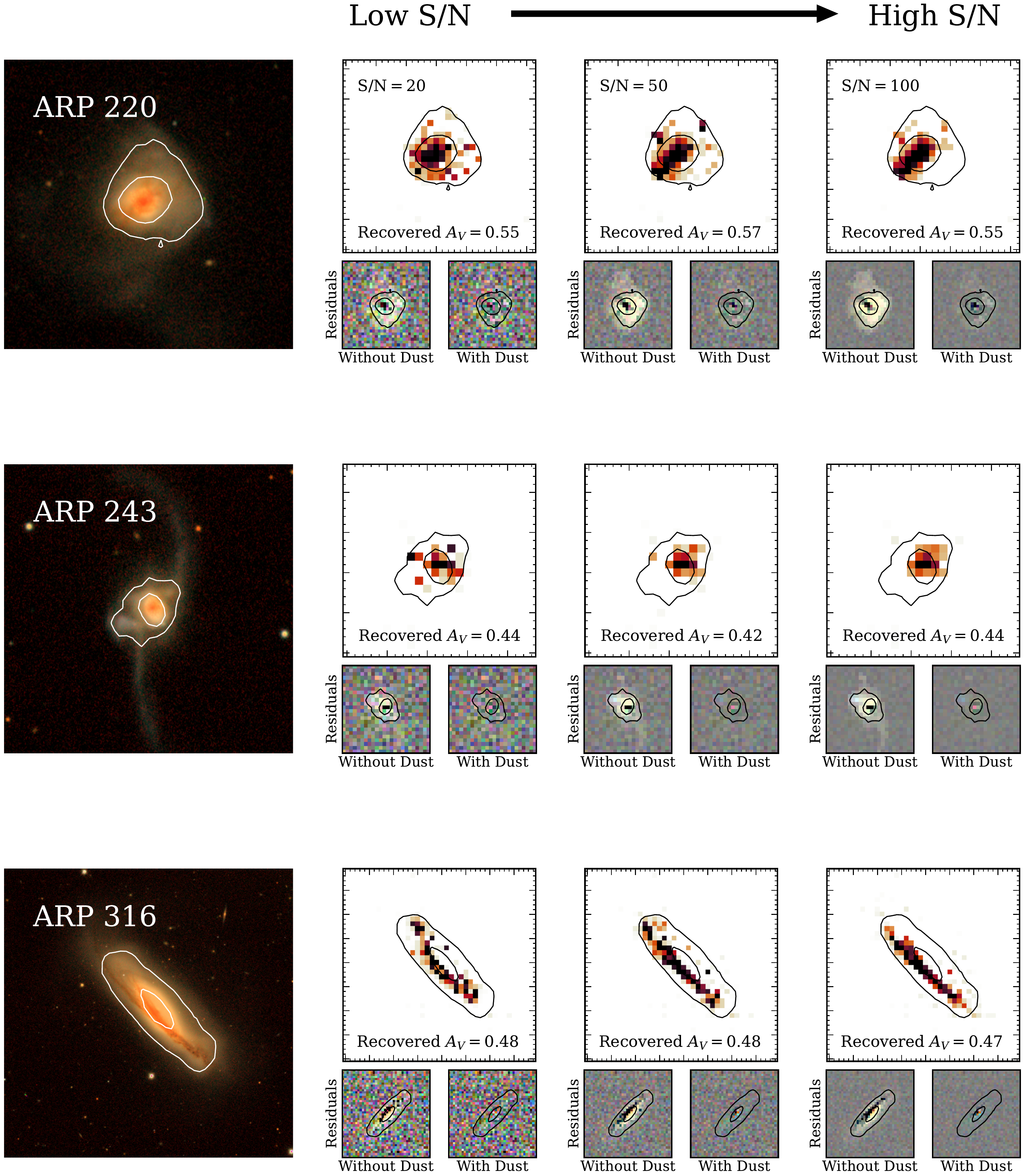}{
\textwidth}{ } }
\caption{
Our joint galaxy--dust modeling applied to three galaxies observed in SDSS DR18.
To represent varied observing conditions, we test three per-band signal-to-noise levels and present the recovered dust map for each galaxy.
We also show the multi-band residuals from fits with and without a dust component. 
The color scale covers per-pixel $A_V\in[0, 1.25]$.
}
\label{fig:arp_summary}
\end{figure*}

Until now, we have assumed the galaxies' unreddened morphologies are known.
This assumption is valid if we have sufficiently red observations; we refer to this scenario as the \textit{red start}.
At such long wavelengths, the galaxies' observed morphologies are only marginally different from their intrinsic (i.e., unreddened) morphologies.
If observations only cover bands significantly affected by dust attenuation, initializing our joint galaxy--dust models is nontrivial; we refer to this scenario as the \textit{blue start}.
As outlined in \autoref{sec:optimization}, we apply a simple method of approximating a galaxy's unreddened morphology from reddened images: assuming the galaxy's bluest pixels are unreddened, we calculate the level of reddening in each of the galaxy's reddest pixels.
We use these dust maps to estimate a galaxy's unreddened morphology and initialize our joint galaxy--dust model.
To avoid the thin-screen degeneracy, the models are still initialized with near zero dust attenuation (see \autoref{sec:optimization}).
Since our method of approximating a galaxy's unreddened morphology is unconstrained (i.e., we enforce no prior or parametric model), the results are naturally noisy.
We therefore rely on our data-driven priors to transform our noisy initializations into plausible galaxies.

We remodel all 65 galaxies from \cite{Faucher2023} under the blue start scenario.
The fit results are shown in the bottom panels of \autoref{fig:dust_fit_summary_breakdown}.
Even with noisy initializations, we accurately recover the galaxy and dust properties for the majority of galaxies (up to the thin-screen degeneracy). 
But compared to our models with optimal initialization, there is somewhat greater scatter in the model accuracy. 
Nonetheless, with the aid of our data-driven priors, we consistently recover galaxy and dust properties from multi-band images.

\subsection{SDSS galaxies}

We apply our joint galaxy--dust model to three galaxies observed by the Sloan Digital Sky Survey \citep{York2000}.
ARP 220, 243, and 316 are local galaxies with clear signs of dust attenuation; all three galaxies are bright ($m_g>16$~mag) and large ($>1$~arcmin across).
In addition to their notoriety as dusty galaxies \citep[all three are included in the Atlas of Peculiar Galaxies,][]{Halton1966}, these galaxies are relatively isochromatic, except for their dust features.

Each galaxy's images are downsized to $64 \times 64$~pixels, and we consider per-band $S/N \in 
[20, 50, 100]$ by injection of Gaussian noise.
We follow the model initialization scheme outlined in \autoref{sec:methods}.
The results of our joint galaxy--dust modeling are presented in \autoref{fig:arp_summary}.
We recover each galaxy's prominent dust feature and estimate the unreddened SED.
By injecting increasing amounts of measurement noise, we find our results are stable across a range of $S/N$ levels.
For context, we also fit each galaxy without dust.
The inclusion of a dust component significantly reduces the multi-band residuals.

\newpage
\section{Discussion}
\label{sec:disc}

Decomposing an observed galaxy into spatially resolved starlight and dust reddening is a highly under-constrained problem.
The likelihood alone cannot distinguish between physical and non-physical galaxy and dust morphologies.
For instance, the attenuation level in regions with little starlight has no effect on the likelihood.
While assuming a parametric form for the galaxy and dust morphologies (e.g., {\sersic}) would guarantee plausible sources, parametric models cannot account for the diversity of galaxy morphologies.
In this work, we model galaxies using non-parametric host and dust morphologies and ensure physical solutions with two data-driven priors: one for the galaxy's intrinsic shape and another for the dust morphology that is conditioned on the galaxy's shape. 

With the aid of our data-driven priors, our model performs well for a variety of galaxy shapes and reddening amplitudes (cf. \autoref{fig:dust_fit_summary_breakdown}).
However, our current implementation still has weaknesses. 
In particular, the likelihood alone cannot distinguish between a galaxy being uniformly dust reddened and the galaxy having an intrinsically red spectrum.
We therefore require the addition of more observables (e.g., spectra and IR imaging) and/or regularization of the galaxy spectrum (e.g., representing the host SED with physically motivated spectral templates).

Optical spectra can constrain the level of reddening within a galaxy via the Balmer decrement: 
comparing the observed ratio of H$\alpha$ to H$\beta$ flux with the expected (dust-free) ratio.
The Balmer decrement measures the attenuation of ionized regions within a galaxy, which is not necessarily the same level of reddening as the rest of the galaxy; however, this measurement is still highly informative.
The presence of dust can also be inferred via FIR dust emission ($3 \lesssim \lambda \lesssim 10^3$~{\microns}).
Unlike the Balmer decrement, FIR emission does not directly measure the level of reddening. 
FIR dust emission is most directly mapped to dust mass \citep[e.g.,][]{Greve2012}, which can be converted to a reddening estimate assuming a geometry for the galaxy and a column density to reddening conversion \citep[e.g.,][]{Guver2009}.
By simultaneously modeling these dust tracers alongside multi-band images, we can break the thin-screen degeneracy and apply our joint galaxy--dust model to arbitrarily dusty galaxies.

The thin-screen degeneracy can further be suppressed by ensuring the recovered shape of the host spectrum is physical. 
To isolate our data-driven morphology priors, we applied minimal regularization to the galaxy spectrum during model fitting.
However, there are key advantages to instead modeling the spectrum with stellar population synthesis methods.
Such an approach will ensure that the recovered spectra are plausible and allows for the galaxy redshifts to be inferred alongside the host and dust properties.

\section{Conclusions}
\label{sec:conclusions}

Outside of the Milky Way, the spatial distribution of interstellar dust has only been measured in a small number of local galaxies with exquisite wavelength coverage and/or spatially resolved spectroscopy \citep[e.g.,][]{Bundy2015,Davies2017}.
These measurements are highly valuable; 
a galaxy's dust morphology encodes the physics of dust production, destruction, and transport. 
Additionally, inference from stellar population synthesis is found to be biased without accurate dust maps \citep[][]{Hahn2024}.
However, for the billions of galaxies soon to be observed by LSST, Euclid, and Roman, we will only have optical and NIR imaging.
In this work, we consider how to simultaneously recover the starlight and dust properties of a galaxy from multi-band images.

Separating the contributions of stars and dust in images is a highly degenerate problem. 
Further complicating matters, galaxy and dust morphologies span a wide variety of complex geometries, limiting the effectiveness of parametric models (e.g., {\sersic}). 
We therefore model galaxies using non-parametric morphologies and encourage physical solutions with the aid of two data-driven priors: the first informs our estimate of the galaxy's intrinsic shape and the second constrains the dust morphology, conditioned on the galaxy's shape. Both of these priors are implemented as score-matching neural networks. 

We test our joint galaxy--dust modeling on both simulated and real galaxies.
Across a wide range of attenuation levels and dust morphologies, we accurately recover the galaxies' host and dust properties.
Our current approach is limited by the thin-screen degeneracy: when all portions of a galaxy are reddened, the likelihood alone cannot separate the contributions of dust and starlight. 
However, treating the host spectrum with a stellar population synthesis model and/or incorporating additional dust tracers (e.g., Balmer decrement and FIR emission) are viable solutions to break this degeneracy.

We have taken the first step towards accurately recovering galaxy and dust properties from multi-band images.
With further development, we anticipate dust and host morphologies can be measured for galaxies at cosmological distances, using only data products available from large upcoming surveys.

\acknowledgments

We thank Andrew Saydjari and 
Bruce Draine for helpful conversations.
JS acknowledges support by the National Science Foundation Graduate Research Fellowship Program under Grant DGE-2039656. 
Any opinions, findings, and conclusions or recommendations expressed in this material are those of the author(s) and do not necessarily reflect the views of the National Science Foundation.

\software{
\scarlettwo{} (Melchior et al., in prep.),  
\texttt{jax} \citep{jax2018},
\texttt{equinox} \citep{equinox2021}
}

\bibliography{paper}%

\end{document}